\documentclass[floatfix,showpacs,pre,aps,a4paper,twocolumn]{revtex4}

 \usepackage{bm}
 \usepackage{amsmath}
 \usepackage{amssymb}
 \usepackage{dsfont}
 \usepackage{latexsym}
 \usepackage{amsfonts}
 \usepackage{epsfig}
 \usepackage{subfigure}
 \usepackage{color}
 \definecolor{darkblue}{rgb}{0,0,.5}
 \usepackage[linktocpage, colorlinks=true, linkcolor=darkblue, citecolor=darkblue]{hyperref}
 \usepackage[all]{hypcap}

\newcommand{\C}[1]{{\cal{#1}}}
\newcommand{\bb}[1]{\textbf{#1}}

\newcommand{\ov}[1]{{\overline{#1}}}

\newcommand{\rrangle}[0]{{\rangle\rangle}}
\newcommand{\llangle}[0]{{\langle\langle}}
\newcommand{\ra}[0]{{\rightarrow}}

\begin{document}

\title{Thermodynamics of Quantum-Jump-Conditioned Feedback Control}

\author{Philipp Strasberg$^1$}
\author{Gernot Schaller$^1$}
\author{Tobias Brandes$^1$}
\author{Massimiliano Esposito$^2$}
\affiliation{
$^1$ Institut f\"ur Theoretische Physik, Technische Universit\"at Berlin, Hardenbergstr. 36, D-10623 Berlin, Germany\\
$^2$ Complex Systems and Statistical Mechanics, University of Luxembourg, L-1511 Luxembourg, Luxembourg}

\begin{abstract}
We consider open quantum systems weakly coupled to thermal reservoirs and subjected to quantum feedback operations 
triggered with or without delay by monitored quantum jumps. We establish a thermodynamic description 
of such system and analyze how the first and second law of thermodynamics are modified by the feedback. 
We apply our formalism to study the efficiency of a qubit subjected to a quantum feedback control 
and operating as a heat pump between two reservoirs. We also demonstrate that quantum feedbacks 
can be used to stabilize coherences in nonequilibrium stationary states which in some cases may 
even become pure quantum states. 
\end{abstract}


\maketitle

\section{Introduction}

Manipulating small quantum systems interacting with their environments with the help of quantum feedback control 
is of crucial importance for modern nanotechnologies. An important step in this direction is to understand 
the thermodynamics of such processes in particular if one wants to use small nanostructures as refrigerators, 
heat pumps or power sources.  

The regime of weak coupling between a system and its environment is nowadays well understood \cite{Breuer02}. 
The system dynamics is described by a Markovian quantum master equation (QME) which in the rotating wave 
approximation (RWA) gives rise to a closed rate equation for the population dynamics in the system eigenbasis 
with rates satisfying local detailed balance. The theoretical framework of stochastic thermodynamics 
can therefore be straightforwardly applied and provides a consistent nonequilibrium thermodynamic 
description for these systems \cite{EspoVdB10_Da}. Without the RWA, QMEs have a more complicated 
structure. 
Although they can be formally written as an ordinary rate equations, the interpretation 
of thermodynamic quantities such as heat becomes ambiguous \cite{EspositoMukamel06}. 

In this paper, we consider open quantum systems subjected to a quantum feedback control 
and described by QMEs in the RWA where the notion of heat remains unambiguously defined. Classical 
and quantum feedback control opens a lot of exciting possibilities to control small systems. For 
instance, feedback can be used to transport electrons against a potential bias \cite{Schaller2011, 
PekolaAverinPRB11, EspositoStrassSchall12} or to control Brownian particles in potential traps 
\cite{Bechhoefer12, SeifertAbreu11, SagawaSano}. From a thermodynamic point of view, feedbacks may inject 
energy as well as entropy into the system and may thus modify the first as well as the second law of 
thermodynamics. Two interesting particular cases are the mechanical work source, which injects energy but 
no entropy, and the Maxwell Demon feedback, which injects entropy but no energy \cite{EspositoSchaller12}. 
Many recent studies have analyzed and quantified the thermodynamic effect of gathering information 
by a measurement performed at pre-determined times and operating back on the system with a time-dependent 
force which depends on the measurement output \cite{SagawaUedaPRL08, SagawaUedaPRL10, SagawaUedaPRE12, 
HorowitzParrondo11, ParrondoHoroSag12, SeifertAbreuPRL12, Rosinberg12}. These setups should be contrasted 
from feedback schemes which rely on a continuous measurement of the system and which operate whenever 
a given signal is detected from the system. The thermodynamics of such feedbacks has been less studied 
and has only been recently considered in Refs. \cite{EspositoSchaller12, EspositoStrassSchall12}.

Our main objective in this study is to characterize the thermodynamic implications of a class of 
feedbacks which were initially introduced in quantum optics by Wiseman and Milburn \cite{WisemanMilburnB}. 
A specific energy or matter transfer between a quantum system and its reservoirs is continuously monitored.
Whenever the transfer event is detected a quantum operation is triggered on the system. The theoretical 
description of such feedback relies on the identification of jump operators in a QME which are associated 
with detection events, e.g. photon emission \cite{WisemanMilburnB, WisemanPRA94}. Such feedbacks can be 
used to stabilize pure quantum states (e.g. of qubits or of the photon field) as was shown theoretically 
\cite{WisemanWang01} as well as experimentally \cite{HarocheNat11, DiCarlo12, Huard13}. 
Similar results were also obtained in quantum transport where jump operators correspond 
to electron detection events \cite{BrandesPoltl11, BrandesKiesslich12}. 

This paper is organized as follows.
In section \ref{ST} we start by reviewing stochastic thermodynamics for open quantum systems 
in general and on a model of a qubit weakly coupled to two thermal reservoirs. 
In section \ref{WMfeed} we introduce the feedback scheme and analyze its effect on the first and second law. 
We also discuss the effect of delays in the feedback and illustrate our results on the qubit model. 
In section \ref{sec example 1} we focus on two applications: a quantum heat pump and the 
stabilization of quantum mechanical coherences. Conclusions are drawn in section \ref{conc}. 

\section{Stochastic thermodynamics of open quantum systems}\label{ST}

We consider a small system with $M$ different energy eigenstates (for instance a quantum dot, a qubit, a molecule 
or a spin), which is described by the system Hamiltonian $H_S = \sum_{i=1}^M E_i|i\rangle\langle i|$ and which is 
weakly coupled to $N$ ideal thermal reservoirs at a given inverse temperature $\beta_\nu =1/T_\nu$ ($k_B\equiv1$ 
throughout the paper) where $\nu \in\{1,\dots,N\}$ is used as an index for the different reservoirs. 
We assume that the reservoirs do not interact directly and solely focus on energy transfer in this paper. 
For simplicity we exclude matter transfer which would require to include the chemical potentials 
of the reservoirs $\mu_\nu$ in the description. 
We also assume that every transition between the energy eigenstates of the system is triggered 
by the absorption/emission of an energy quantum from/to a particular reservoir $\nu$. 

\begin{figure}[h]
\includegraphics[width=0.28\textwidth,clip=true]{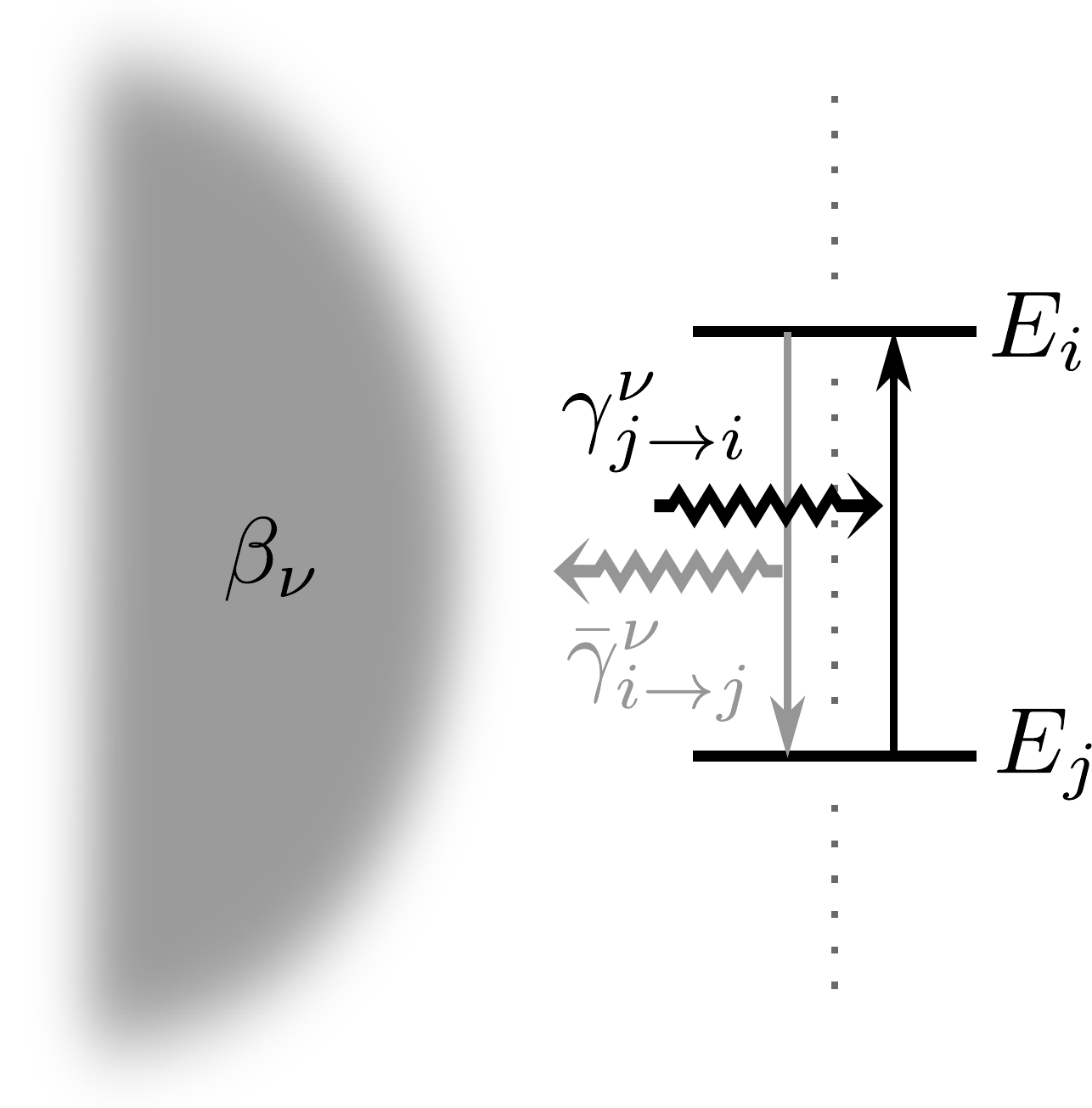}
\caption{\label{fig general setup} Illustration of a possible transition between two energy levels $E_i$ and $E_j$ 
of the system by absorption of an energy quantum (in black) or an emission of an energy quantum (in gray) from or into 
the thermal reservoir $\nu$.}
\end{figure}

\subsection{Dynamics}

In the weak coupling regime where the Born-Markov approximation is justified, the time evolution 
of the system density matrix $\rho$ follows the Markovian QME \cite{Breuer02}:
\begin{equation}\label{eq generic QME}
 \frac{\partial}{\partial t}\rho(t) = \C W \rho(t).
\end{equation}
The generator $\C W$ is a superoperator/operator in Hilbert/Liouville space. 
To fix the notation we use latin letters $i,j,k$ to label the energy levels of the 
system and $|i\rangle,|j\rangle,|k\rangle$ are the energy eigenstates in the system Hilbert space. 
In Liouville space we order the elements of the density matrix into the vector $(\rho_{pop},\rho_{coh})$ 
where $\rho_{pop} = (p_1,\dots,p_M)$ describes the populations $p_i = \langle i|\rho|i\rangle$ and 
$\rho_{coh} = (\rho_{12},\rho_{21},\dots,\rho_{M-1M},\rho_{MM-1})$ is the vector of the $M(M-1)$ 
coherences $\rho_{ij} = \langle i|\rho|j\rangle, i\neq j$. Consequently, the trace of a matrix 
becomes a sum over the first $M$ entries of the corresponding vector in Liouville space. Furthermore, we order the 
indexes $\{i\}$ such that $i>j$ implies also $E_i>E_j$. Thus, we exclude the possibility of degeneracies 
in the system. To use an intuitive notation we will also use the indexes $i,j,k$ to refer to the 
populations in Liouville space, i.e., whenever there is a sum over $i,j,k$ it runs from $1$ to $M$. 
The corresponding states in Liouville space are denoted by $|i\rrangle,|j\rrangle,|k\rrangle$. 
We will not introduce a notation for the coherences because we do not need them. 

In the energy eigenbasis of the system Hamiltonian, due to the RWA the generator has a block structure of the form 
\begin{equation}\label{eq block structure generator}
 \C W  =  \left(\begin{array}{cc}
               \C W_{pop}	&	0		\\
               0		&	\C W_{coh}	\\
           \end{array}\right),
\end{equation}
which shows that population and coherences evolve independently from each other in that basis.
In fact the populations obey a rate equation $\partial_t\rho_{pop} = \C W_{pop}\rho_{pop}$, 
while the coherences are exponentially damped and vanish at steady-state. 
The generator of the populations can be made more explicit   
\begin{equation}\label{eq generator with jumpers}
 \C W_{pop} = \C W_0 + \sum_{\nu}\sum_{i>j}\big(\C J_{j\rightarrow i}^\nu+\ov{\C J}_{i\rightarrow j}^\nu\big),
\end{equation}
where $\C J_{j\rightarrow i}^\nu$ ($\ov{\C J}_{i\rightarrow j}^\nu$) is a jump-superoperator responsible 
for a quantum jump upwards from level $j$ to $i$ (downwards from level $i$ to $j$) corresponding to an energy 
$E_i-E_j>0$ entering (exiting) the system from (to) reservoir $\nu$ as illustrated on Fig. \ref{fig general setup}. 
Mathematically, these jump operators can be expressed as 
\begin{equation}\label{eq jump operators}
 \C J_{j\rightarrow i}^\nu = \gamma_{j\rightarrow i}^\nu|i\rrangle\llangle j|, ~~~ \ov{\C J}_{i\rightarrow j}^\nu = \ov\gamma_{i\rightarrow j}^\nu|j\rrangle\llangle i|
\end{equation}
where $\gamma_{j\rightarrow i}^\nu$ and $\ov\gamma_{i\rightarrow j}^\nu$ are rates satisfying local detailed balance: 
\begin{equation}\label{eq local detailed balance}
 \ln(\gamma_{j\rightarrow i}^\nu/\ov\gamma_{i\rightarrow j}^\nu) = -\beta_\nu(E_i-E_j).
\end{equation}
If we have correctly identified all the jumps the remaining part of the generator $\C W_0$ has the form 
\begin{equation}\label{eq W0 explicit}
 \C W_0 = -\sum_{\nu}\sum_{i>j} \big(\gamma_{j\rightarrow i}^\nu|j\rrangle\llangle j| + \ov\gamma_{i\rightarrow j}^\nu|i\rrangle\llangle i|\big).
\end{equation}

The probability current associated to the transition $(\nu,(i,j))$ is given by (see appendix \ref{sec full counting statistics})
\begin{equation}\label{eq steady state current formula}
 I_{(i,j)}^\nu(t) = \gamma_{j\rightarrow i}^\nu p_j(t) - \ov\gamma_{i\rightarrow j}^\nu p_i(t).
\end{equation}
Often however, we are only interested in the long time steady state behavior of the system where the probabilities $p_j(t)$ 
become time-independent and fulfill $\sum_j (\C W_{pop})_{ij} p_j = 0$ for all $i$. In this case, we adopt the 
simple notation $p_j = \lim_{t\rightarrow\infty} p_j(t), I_{(i,j)}^\nu = \lim_{t\rightarrow\infty} I_{(i,j)}^\nu(t)$ 
etc., i.e., we drop the time dependence if we talk about the steady state. 

\subsection{Thermodynamics}

Since we have identified the currents for each transition, we can now introduce the heat flow from reservoir $\nu$ as 
\begin{equation}\label{eq heat flow in general}
 \dot Q^{(\nu)}(t) = \sum_{i>j} (E_i-E_j) I_{(i,j)}^\nu(t).
\end{equation}
It is by definition positive if it enters the system. We denote the change in the system energy by 
\begin{equation}
 \dot E(t) \equiv \frac{d}{dt}\mbox{tr}[\C H_S\rho] = \sum_i E_i \dot p_i(t)
\end{equation}
where $\C H_S$ is a formal expression of the Hamiltonian in Liouville space with entries $\C H_S = \sum_i E_i|i\rrangle\llangle i|$. 
The \emph{first law of thermodynamics} now demands that the change in system energy is balanced by the heat flows into the system 
(see appendix \ref{sec derivation first law})
\begin{equation}\label{eq first law}
 \dot E(t) = \sum_\nu \dot Q^{(\nu)}(t).
\end{equation}
At steady state, the left hand side vanishes and the first law becomes $\sum_\nu \dot Q^{(\nu)}=0$. 

We will use the Shannon entropy to characterize the entropy of the system. We note that due to the structure 
of Eq. (\ref{eq block structure generator}), at steady state the von Neumann entropy of the system coincides with 
the Shannon entropy of the system. As customary in stochastic thermodynamics, the change in Shannon entropy 
of the system can be split in two parts \cite{EspoVdB10_Da}: 
\begin{equation}\label{eq entropy splitting}
 \frac{d}{dt}S(t) = -\frac{d}{dt}\sum_i p_i(t)\ln p_i(t) = \dot S_{\bf i}(t) + \dot S_{\bf e}(t),
\end{equation}
where $\dot S_{\bf i}(t)$ is the non-negative entropy production which quantifies the irreversibility of the dynamics 
and $\dot S_{\bf e}(t)$ is the entropy flow arising from heat exchanges with the environments. More explicitly
\begin{align}\label{eq entropy prod no feedback}
 \dot S_{\bf i}(t) &= \sum_{\nu} \sum_{i,j} \C W_{ij}^{(\nu)}p_{j}(t)\ln\frac{\C W_{ij}^{(\nu)}p_{j}(t)}{\C W_{ji}^{(\nu)}p_{i}(t)} \ge 0, \\
 \dot S_{\bf e}(t) &= \sum_{\nu}\sum_{i,j} \C W_{ij}^{(\nu)}p_{j}(t)\ln\frac{\C W_{ji}^{(\nu)}}{\C W_{ij}^{(\nu)}},
\end{align}
where $\C W_{ij}^{(\nu)}$ denotes the matrix elements of the population generator (\ref{eq generator with jumpers})
associated to transitions triggered by the reservoir $\nu$. 

Using the local detailed balance relation in Eq. (\ref{eq local detailed balance}), it is easy to show 
that the entropy flow can be expressed as (minus) the reversible entropy changes in the reservoirs 
$\dot S_{\bf e}(t) = \sum_\nu \dot Q^{(\nu)}(t)/T_\nu$. Hence, 
\begin{equation}\label{eq second law no feedback}
 \dot S_{\bf i}(t) = \dot S(t) - \sum_\nu \frac{\dot Q^{(\nu)}(t)}{T_\nu} \ge 0,
\end{equation}
which corresponds to the \emph{second law of thermodynamics}. 
At steady state the change in system entropy vanishes, $\dot S = 0$, such that $\dot S_{\bf i} = -\dot S_{\bf e}$.

\subsection{Qubit model}\label{secEx1}

As our paradigmatic model we consider a qubit weakly coupled to two bosonic reservoirs $L$ and $R$. 
Such spin-boson-models have been studied extensively in the literature for instance to understand heat 
pumps and thermal transport through molecules \cite{SegalNitzan06, SegalPRL08, HanggiBaowenPRL10}. 
We consider phonon reservoirs, but photons could be considered as well. 
The total Hamiltonian is the sum of three contributions
\begin{align}
 H	&=	H_S + H_B + V, \\
 H_S	&=	\frac{\Omega}{2}(|1\rangle\langle1|-|0\rangle\langle0|),	\\
 H_B	&=	\sum_{\nu\in\{L,R\}} \sum_{\bb q} \omega_{\bb q\nu} b_{\bb q\nu}^\dagger b_{\bb q\nu},	\\
 V	&=	\sum_{\nu\in\{L,R\}} \sum_{\bb q} T_{\bb q\nu} (b_{\bb q\nu}^\dagger |0\rangle\langle1| + b_{\bb q\nu} |1\rangle\langle0|),
\end{align}
where $\omega_{\bb q\nu} >0$, $\Omega>0$, $T_{\bb q\nu} \in\mathbb{R}$, and $b_{\bb q\nu}$ are 
bosonic annihilation operators. We considered the interaction in the RWA. 

The master equation of this system in the Born-Markov approximation is well-known \cite{Breuer02, WisemanMilburnB}. 
In the basis $(p_0 = \langle0|\rho|0\rangle, p_1 = \langle1|\rho|1\rangle, \rho_{01},\rho_{10})$ 
the generator has the structure (\ref{eq block structure generator}) where 
\begin{align}
 \C W_{pop}	&=	\sum_\nu\left(\begin{array}{cc}
           	  	       -\gamma_\nu	&	\ov\gamma_\nu	\\
           	  	       \gamma_\nu	&	-\ov\gamma_\nu	\\
           	  	      \end{array}\right),	\\
 \C W_{coh}	&=	\left(\begin{array}{cc}
           	  	       i\Omega'-\sum_\nu\frac{\gamma_\nu+\ov\gamma_\nu}{2}	&	0	\\
           	  	       0	&	-i\Omega'-\sum_\nu\frac{\gamma_\nu+\ov\gamma_\nu}{2}	\\
           	  	      \end{array}\right)	\label{eq Wcoh qubit}
\end{align}
The rates $\gamma_\nu,\ov\gamma_\nu$ are determined by the Bose distribution 
$n_\nu(\Omega) = (e^{\beta_\nu\Omega}-1)^{-1}$ evaluated at the level-splitting: 
$\gamma_\nu = \Gamma_\nu n_\nu(\Omega), \ov\gamma_\nu = \Gamma_\nu(1+n_\nu(\Omega))$ with $\Gamma_\nu>0$. 
$\Omega'$ is the renormalized level-splitting due to the Lamb shifts. 

We now focus on the steady state behavior of the system. 
In this $p_0 =1-p_1= \big(\sum_\nu \ov \gamma_\nu\big)/\big(\sum_\nu(\gamma_\nu+\ov\gamma_\nu)\big)$ and the coherences 
vanish $\rho_{01} = \rho_{10} = 0$. The probability current (\ref{eq steady state current formula}) is
given by $I^\nu = \gamma_\nu p_0 - \ov\gamma_\nu p_1$ and the corresponding heat flow becomes 
$\dot Q^{(\nu)} = \Omega I^\nu$. Consequently, the first and second law of thermodynamics read 
\begin{equation}\label{eq second law}
 \text{I.} ~ \dot Q^L + \dot Q^R = 0, ~~~ \text{II.} ~ (\beta_R-\beta_L) \dot Q^L \ge 0.
\end{equation}
The second law expresses the fact that on average the phonons are flowing from the hot to the cold reservoir. 

\section{Thermodynamics of Wiseman-Milburn feedbacks}\label{WMfeed}

A feedback describes the situation in which a system is measured and according to the measurement 
output a certain operation is performed on it. In our case the identification of jump processes 
defines a weak measurement of the system by the reservoirs. Note that also ``no signal'', i.e., the time between two subsequent jumps, 
reveals information about the system. The idea of Wiseman and Milburn was now to use the signal of detection events 
to trigger control operations on the system \cite{WisemanMilburnB, WisemanPRA94}. 

\subsection{Control Operations}

To describe the situation with feedback we introduce control superoperators $\C C_{j\rightarrow i}^\nu$ 
($\ov{\C C}_{i\rightarrow j}^\nu$), which act on the system after a certain absorption (emission) process 
in the system has been induced by the reservoir $\nu$. Most of the time we will assume that the control 
operation acts instantaneously after a jump on the system, i.e., the feedback is much faster than all 
other relevant time scales of the system (the case of a finite delay will be treated in Sec. \ref{sec delay}). 
The resulting effective generator can be written as \cite{WisemanMilburnB, WisemanPRA94} 
\begin{equation}\label{eq generator feedback}
 \C W^C = \C W_0 + \sum_{\nu}\sum_{i>j} \big(\C C_{j\rightarrow i}^\nu\C 
J_{j\rightarrow i}^\nu + \ov{\C C}_{i\rightarrow j}^\nu\ov{\C J}_{i\rightarrow j}^\nu\big).
\end{equation}
Using the form of the jump operators, Eq. (\ref{eq jump operators}), we see that this generator has a block structure of the form 
\begin{equation}\label{eq block structure with feedback}
 \C W^C = \left(\begin{array}{cc}
                 \C W_{pop}^C	&	0		\\
                 \C W_{cp}^C	&	\C W_{coh}	\\
                \end{array}\right).
\end{equation}
The populations still evolve independently from the coherences, but the coherences 
get affected by the populations. This implies that such a feedback is able to build up 
coherences in the steady state, as we will see in detail in section \ref{sec stabilization}. 

The quantum control operation is chosen as a unitary operation $U_{\C C}$ ($U_{\C C}^\dagger U_{\C C}=1$) 
in the Hilbert space of the system: $\C C \rho \leftrightarrow U_{\C C} \varrho U_{\C C}^\dagger$, 
where $\varrho$ denotes the density matrix acting in Hilbert space whereas $\rho$ denotes the corresponding 
vector in Liouville space. 
Defining the transition probability due to the control operation 
$\C C_{ki} \equiv \llangle k|\C C|i\rrangle = \vert \langle k| U_{\C C} |i\rangle \vert^2$ from population $i$ to population $k$, we find that
\begin{equation}\label{eq useful formula}
 \sum_{k=1}^M \C C_{ki} = \sum_{i=1}^M \C C_{ki} = 1 .
\end{equation}

The effective population generator can be explicitly written as
\begin{equation}
 \begin{split}\label{eq pop generator with feedback}
  \C W_{pop}^C	&=	\sum_{\nu}\sum_{i>j} \gamma_{j\rightarrow i}^\nu\left(\sum_k (\C C_{j\rightarrow i}^\nu)_{ki}|k\rrangle\llangle j| - |j\rrangle\llangle j|\right)	\\
		&+	\sum_{\nu}\sum_{i>j} \ov\gamma_{i\rightarrow j}^\nu\left(\sum_k (\ov{\C C}_{i\rightarrow j}^\nu)_{kj}|k\rrangle\llangle i| - |i\rrangle\llangle i|\right).
 \end{split}
\end{equation}
The dynamics of the populations can be interpreted as follows. Immediately after the detection of a 
jump from $|j\rrangle\rightarrow|i\rrangle$, the control operation generates a further ``jump'' from
$|i\rrangle\rightarrow|k\rrangle$ with probability $(\C C_{j\rightarrow i}^\nu)_{ki}$. 
In absence of feedback $({\C C}_{j \rightarrow i}^\nu)_{ki} = \delta_{ki}$ and we recover the generator 
(\ref{eq generator with jumpers}). If the feedback operation commutes with the system Hamiltonian, 
$[C,H_S] = 0$, the generator remains unaffected by the control operation: $\C W^C = \C W$. 

\subsection{Thermodynamics of feedback}

The probability current associated with the $\nu$-th reservoir induced transition 
$|j\rrangle\rightarrow|i\rrangle$ is given by 
\begin{equation}\label{eq probability flow with feedback}
 I_{(i,j)}^\nu(t) = \gamma_{j\rightarrow i}^\nu p_j(t) - \ov\gamma_{i\rightarrow j}^\nu p_i(t) .
\end{equation}
This result is derived using counting statistics methods in appendix \ref{sec full counting statistics}. 
Not surprisingly, we get the same expression as without feedback, Eq. (\ref{eq steady state current formula}), 
but where the steady state probabilities are obtained from the generator $\C W^C$ instead of $\C W$. 
The heat flow is consequently given by 
\begin{equation} \label{HeatFeed}
\dot Q^{(\nu)}(t) = \sum_{i>j} (E_i-E_j) I_{(i,j)}^\nu(t)
\end{equation}
as in Eq. (\ref{eq heat flow in general}). 

The rate of energy injection by the control operation after a transition $|j\rrangle\rightarrow|i\rrangle$ triggered by reservoir $\nu$ 
can be obtained from the difference between the energy of the system after the transition and the subsequent control operation 
and the energy of the system right after the transition but before the control operation: 
\begin{equation}\label{eq work feedback general}
 \begin{split}
  (\dot{\C F}^{(\nu)}_E)_{(i,j)}(t) &=	\mbox{tr}[\C H_S(\C C_{j\rightarrow i}^\nu-1)\C J_{j\rightarrow i}^\nu \rho(t)]	\\
				    &+	\mbox{tr}[\C H_S(\ov{\C C}_{i\rightarrow j}^\nu-1)\ov{\C J}_{i\rightarrow j}^\nu\rho(t)].
 \end{split}
\end{equation}
It is positive if the feedback in average \emph{injects} energy into the system and it is 
zero in absence of feedback (i.e., the $\C C_{i\rightarrow j}^\nu$ are the identity operator). 

The change in the system energy has now to be balanced by the heat flows (\ref{HeatFeed}) \emph{and} the energy injected by the 
feedback (\ref{eq work feedback general}). The first law of thermodynamics thus gets modified according to 
\begin{equation}\label{eq first law with feedback}
 \dot E(t) = \sum_\nu \left(\dot Q^{(\nu)}(t) + \dot{\C F}_E^{(\nu)}(t)\right)
\end{equation}
where $\dot{\C F}_E^{(\nu)}(t) \equiv \sum_{i>j} (\dot{\C F}_E^{(\nu)})_{(i,j)}(t)$. 
This result is explicitly derived in appendix \ref{sec derivation first law}. 

\begin{figure}[h]
\includegraphics[width=0.29\textwidth,clip=true]{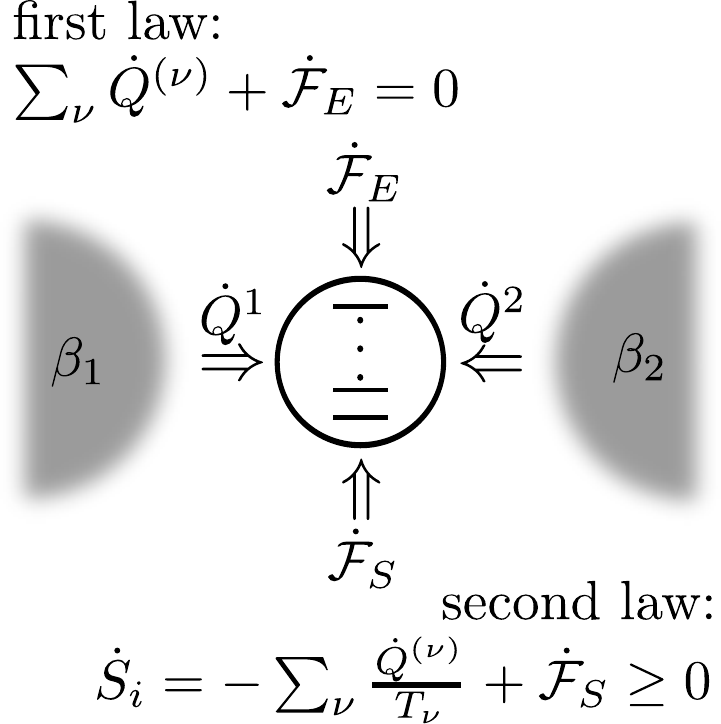}
\caption{\label{fig thermodynamics} Summary of the modified first and second law for a system at steady state subjected to 
feedback control in contact with two reservoirs. Both, the first and second law are modified due to the rate of energy injection 
$\dot{\C F}_E$ by the feedback and the information flow $\dot{\C F}_S$.}
\end{figure}

We now turn to the second law. Since the population generator is still a proper rate matrix,
the change in time of the Shannon entropy can be again splitted as in Eq. (\ref{eq entropy splitting}) and 
the entropy production $\dot S_{\bf i}$ and the entropy flow $\dot S_{\bf e}$ are defined as usual as \cite{EspoVdB10_Da} 
\begin{align}
 \dot S_{\bf i}(t) &= \sum_{\nu} \sum_{i,j} (\C W^C)_{ij}^{(\nu)}p_{j}(t)
\ln\frac{(\C W^C)_{ij}^{(\nu)}p_{j}(t)}{(\C W^C)_{ji}^{(\nu)}p_{i}(t)} \ge 0,	\label{eq entropy prod general}	\\
 \dot S_{\bf e}(t) &= \sum_{\nu}\sum_{i,j} (\C W^C)_{ij}^{(\nu)}p_{j}(t)
\ln\frac{(\C W^C)_{ji}^{(\nu)}}{(\C W^C)_{ij}^{(\nu)}}.	\label{eq entropy flow general}
\end{align}
We remark however that, since coherences survive at steady state, the von Neumann entropy 
$\C S = -\mbox{tr}[\rho\ln\rho]$ is not equivalent anymore to the Shannon entropy.
This suggests that (\ref{eq entropy prod general}) is not the only possible choice of entropy production.
An approach based on the splitting of the evolution of the system von Neumann entropy as an entropy 
production and entropy flow term (in the basis diagonalizing the system density matrix) as proposed 
in \cite{EspositoMukamel06} could also have been attempted.
In any case, the choice of entropy production (\ref{eq entropy prod general}) constitutes a non-negative 
quantity which only cancels when all probability currents between pairs of populations due to a transition 
from a reservoir $\nu$ and its corresponding feedback operation vanish: $(\C W^C)_{ij}^{(\nu)}p_{j}
=(\C W^C)_{ji}^{(\nu)}p_{i}$.

This means that the modified entropy flow reads now 
\begin{equation}\label{eq entropy flow feedback}
 \dot S_{\bf e}(t) = \sum_\nu \frac{\dot Q^{(\nu)}(t)}{T_\nu} -\dot{\C F}_S(t)
\end{equation}
where $\dot{\C F_S}(t)$ characterizes the influence of the feedback on the entropy balance (or the ``second 
law of thermodynamics''). Its explicit expression is given in appendix \ref{sec derivation first law}. 
This information flow $\dot{\C F_S}(t)$ will be useful to define notions of feedback efficiency 
as we will see in Sec. \ref{sec efficiency qubit}. 
In the steady state regime we have once again that 
\begin{equation}\label{eq entropy prod feedbackSS}
\dot S_{\bf i} = -\dot S_{\bf e} = - \sum_\nu \frac{\dot Q^{(\nu)}}{T_\nu} + \dot{\C F}_S \geq 0.
\end{equation}

In summary, we have introduced two new quantities $\dot{\C F}_E$ and $\dot{\C F}_S$ to take into account 
the influence of the feedback on the energy balance (the first law) and on the entropy balance (the second 
law) on the system. Both are additive in terms of the reservoirs: $\dot{\C F}_{E,S} = \sum_\nu \dot{\C F}^{(\nu)}_{E,S}$. 
We can distinguish two limiting regimes of feedback control, an energy dominated regime for 
$|\dot{\C F}_E| \gg 0$ and $\dot{\C F}_S \approx 0$ and an entropy (or information) 
dominated feedback for $\dot{\C F}_E \approx 0$ and $|\dot{\C F}_S| \gg 0$. 
The latter corresponds to the class of ``Maxwell demon feedback'' introduced in Ref. \cite{EspositoSchaller12}. 
The Wiseman-Milburn feedbacks presented here can however never fully operate as ``Maxwell demon feedback'' 
since by construction the control operation (if it does not commute with the system Hamiltonian) 
always injects or removes energy from the system. 
This statement is expected not to hold any longer if we consider quantum systems with degenerate states 
because the feedback operation may give rise to non-trivial effects without energy consumption in the 
degenerate subspace. The present thermodynamic analysis is summarized in Fig. \ref{fig thermodynamics}. 

\subsection{Delayed Feedback Control} \label{sec delay}

In this section we briefly discuss how situations with a finite delay between the measurement 
and the control operation can be treated within our framework. A detailed discussion of the 
thermodynamic influence for time-delayed feedback control is however beyond the scope of 
the present paper. 

The theory was developed in Ref. \cite{Emary12} where it was shown that an arbitrary delay 
time still leads to a non-Markovian master equation description provided one performs the 
so-called ``control-skipping assumption''. This assumption demands that the control operation 
is skipped when another jump is monitored during the delay. The resulting master equation reads 
\begin{equation}
 \begin{split}\label{eq master equation with delay}
  \dot\rho(t)	&=	\big(\C W_0 + \sum_\alpha \C J_\alpha\big)\rho(t)	\\
		&+	\sum_\alpha (\C C_\alpha-1)e^{\C W_0\tau_\alpha} \C J_\alpha \theta(t-\tau_\alpha)\rho(t-\tau_\alpha).
 \end{split}
\end{equation}
Here, the set $\{\alpha\}$ enumerates all the possible jumps $\C J_\alpha$ which are followed 
by some control operation $\C C_\alpha$ after a certain delay time $\tau_\alpha$ and 
$\theta(t-\tau_\alpha)$ denotes the Heaviside step function. 

We verify that we recover the master equation (\ref{eq generic QME}) with (\ref{eq generator with jumpers}) 
in absence of feedback (when $\C C_\alpha = 1$ for all $\alpha$) and the master equation (\ref{eq generic QME}) 
with (\ref{eq generator feedback}) in case of vanishing delay time $\tau_\alpha=0$.
For an infinite delay time, using (\ref{eq W0 explicit}) we have that $\lim_{\tau_\alpha\rightarrow\infty}
e^{\C W_0\tau_\alpha} = 0$ and thus we recover our master equation without feedback. This makes sense since 
for an infinite delay time the control operation is never performed due to the ``control-skipping assumption''. 

We would like to define heat, work and entropy for the non-Markovian equation (\ref{eq master equation with delay}). 
However, the theory of stochastic thermodynamics for non-Markovian dynamics requires some care \cite{SeifertSpeck07, 
EspositoLindenberg, BreuerMessinaPRE13} and is not as straightforward as its Markovian counterpart.
We will focus therefore exclusively on the steady state behavior. 

Let us consider Eq. (\ref{eq master equation with delay}) in Laplace space (the Laplace transform of an arbitrary 
function of time $f(t)$ is defined as $\hat f(z) \equiv \int_0^\infty dt e^{-zt} f(t)$). We get \cite{Emary12} 
\begin{equation}
 z\hat\rho(z) - \rho(0) = \C W_{delay}(z)\hat\rho(z)
\end{equation}
where
\begin{equation}\label{eq generator delay}
 \C W_{delay}(z) = \C W_0 + \sum_\alpha\big[1 + (\C C_\alpha-1)e^{(\C W_0-z)\tau_\alpha}\big]\C J_\alpha.
\end{equation}
This system exhibits one (or several) non-trivial steady states 
$\rho \equiv \lim_{z\searrow0}z\hat\rho(z) = \lim_{t\rightarrow\infty}\rho(t)$ 
if $\C W_{delay}(0)$ has one (or several) zero eigenvalues. 
For large $t$ when the system is close to steady state we have $\rho(t-\tau_\alpha) \approx \rho(t)$ and $\theta(t-\tau_\alpha) = 1$ 
for all $\tau_\alpha$ and thus
\begin{equation}\label{eq master equation delay long times}
 0 \approx \frac{\partial}{\partial t}\rho(t) = \C W_{delay}(0)\rho(t),
\end{equation}
where $\C W_{delay}(0)$ is a well-defined Markovian generator. This generator can be interpreted as a feedback 
generator (\ref{eq generator feedback}) without delay if we choose as a control operation $\tilde{\C C}_\alpha 
= 1 + (\C C_\alpha-1)e^{\C W_0 \tau_\alpha}$. This operation still fulfills the condition (\ref{eq useful formula}).
Furthermore, since we still have an additive structure of the form $\C W_{delay} = \sum_\nu \C W_{delay}^{(\nu)}$, 
we can define the heat flow, the energy injected by the feedback, and the entropy production, in the same way 
as before for Markovian dynamics. We will use Eq. (\ref{eq master equation delay long times}) to investigate 
numerically the impact of a time delay on the thermodynamics in Sec. \ref{sec example 1}. 

\subsection{Qubit model}\label{qubitModelFeed}

We reconsider the qubit weakly coupled to two thermal reservoirs presented in Sec. \ref{secEx1}.
Four different types of jumps can occur in this system: the qubit can absorb ($+$) or emit ($-$) 
a phonon from or into the $\nu$'th reservoir. Upon detection of these jumps the qubit is subjected 
to a quantum control operation $\C C^\nu_\pm$ performing a unitary operation 
$\C C^\nu_\pm\rho \leftrightarrow U^\nu_\pm\varrho (U^\nu_\pm)^\dagger$, where 
$U^\nu_\pm\equiv\exp\big[-i\alpha^\nu_\pm(|0\rangle\langle1|+|1\rangle\langle0|)\big]$ 
rotates the qubit around the $x$-axis on the Bloch sphere by an angle $\alpha^\nu_\pm$. 
Therefore, instead of being in the ground or excited state right after the emission or absorption 
of a phonon, the system ends up in a superposition of energy eigenstates due to the control operation. 
The generator with feedback has the structure (\ref{eq block structure with feedback}) with 
\begin{equation}
 \begin{split}\label{eq feedback generator qubit}
  \C W^C_{pop} &= \sum_\nu \left(\begin{array}{cc}
                    -\gamma_\nu\cos^2\alpha_+^\nu	&	\ov\gamma_\nu\cos^2\alpha_-^\nu		\\
                    \gamma_\nu\cos^2\alpha_+^\nu	&	-\ov\gamma_\nu\cos^2\alpha_-^\nu	\\
                   \end{array}\right),	\\
  \C W^C_{cp} &= \frac{i}{2}\sum_\nu \left(\begin{array}{cc}
                    -\gamma_\nu\sin2\alpha_+^\nu	&	\ov\gamma_\nu\sin2\alpha_-^\nu	\\
                    \gamma_\nu\sin2\alpha_+^\nu		&	-\ov\gamma_\nu\sin2\alpha_-^\nu	\\
                   \end{array}\right),
 \end{split}
\end{equation}
whereas $\C W^C_{coh}$ remains unaffected by the feedback and is thus given by (\ref{eq Wcoh qubit}). 

According to Eq. (\ref{eq first law with feedback}), the first law in presence of feedback reads 
\begin{equation}\label{FirstLawModel}
 \dot E(t) = \dot Q^L(t) + \dot{\C F}_E^L(t) + \dot Q^R(t) + \dot{\C F}_E^R(t),
\end{equation}
where the rate of energy injection due to feedback (\ref{eq work feedback general}) is given by
\begin{equation}\label{eq work feedback qubit}
 \dot{\C F}_E^{(\nu)}(t) = \Omega[I^\nu_F(t)-I^\nu(t)] .
\end{equation}
Here, we introduced the effective current $I^\nu_F(t) = \cos^2\alpha_+^\nu\gamma_\nu p_0(t) - \cos^2\alpha_-^\nu\ov\gamma_\nu p_1(t)$. 
At steady state $\dot E = 0$ and the first law can be rewritten as $I^L_F+I^R_F=0$. 
The steady state of the qubit is given for completeness in appendix \ref{sec steady state qubit}. 

At steady state, the second law of thermodynamics in presence of feedback is given by 
(\ref{eq entropy prod feedbackSS}), where the entropy current generated by the feedback reads 
\begin{equation}
 \dot{\C F}_S = \sum_\nu\frac{\dot Q^{(\nu)}}{T_\nu} + \left(-\frac{\Omega}{T_L} + \frac{\Omega}{T_R} + f_L - f_R\right) I_F^L.
\end{equation}
We introduced $f_\nu = \ln\frac{\cos^2\alpha_+^\nu}{\cos^2\alpha_-^\nu}$. 
More explicitly, this means that the entropy production is given by 
\begin{equation}
 \begin{split}\label{eq entropy prod qubit feedback}
  \dot S_{\bf i} 
	   &=	\left(-\frac{\Omega}{T_L} + \frac{\Omega}{T_R} + f_L - f_R\right) I^L_F \ge 0.
 \end{split}
\end{equation}

The stationary regime in presence of a finite time delay can also be considered. 
Using $\C W_{delay}(0)$ from Eq. (\ref{eq generator delay}, \ref{eq master equation delay long times}) 
we can calculate the steady state probabilities $p_0(\tau_\alpha), p_1(\tau_\alpha)$ for arbitrary 
delay times $\tau_\alpha$. They are unique in this model. Using the full counting statistics methods from 
appendix \ref{sec full counting statistics}, we can evaluate the heat flows $\dot Q^{(\nu)}$ from which we 
can directly infer that $\dot{\C F}_E = -\dot Q^L - \dot Q^R$ due to the first law. Furthermore, using Eq. 
(\ref{eq entropy prod general}) to calculate the entropy production, we also easily find 
$\dot{\C F}_S = \dot S_{\bf i} + \dot Q^L/T_L + \dot Q^R/T_R$. 

\section{Applications} \label{sec example 1}

In this section we will focus on two particular applications of the feedback scheme developed above. 
We will first study a quantum controlled heat pump and use our formalism to define its efficiency. 
We will also numerically analyze the effect of finite delay times in the feedback. Then we will 
study how efficient the feedback can stabilize pure quantum states in the qubit. 

\subsection{Heat Pump}
\label{sec efficiency qubit}

A heat pump is a device operating between two thermal reservoirs and using work to deliver heat to 
the hot reservoir. For definiteness we choose $T_L > T_R$. Other thermodynamic engines such as a 
refrigerator or a power source can be treated in a very similar way. 

The efficiency of a conventional heat pump is characterized by the \emph{coefficient of performance} 
which quantifies how much heat can be transfered to the hot reservoir $-\dot Q^L \geq 0$ by using 
a particular amount of work $\dot W>0$:
\begin{equation}
 \kappa \equiv \frac{-\dot Q^L}{\dot W} \le \frac{T_L}{T_L-T_R} \equiv \frac{1}{\eta_C},
\end{equation}
where $\eta_C$ is the Carnot efficiency. It is bounded between zero and the inverse Carnot efficiency.  

In our setup, one would be tempted to replace the conventional work source by the external source 
of energy $\dot{\C F}_E$ injected by the feedback, and thus to define 
\begin{equation}\label{eq cop}
 \tilde\kappa \equiv \frac{-\dot Q_L}{\dot{\C F}_E}.
\end{equation}
While meaningful this quantity is not bounded by the theory and as we will see can become greater 
than $1/\eta_C$. Nevertheless, our formalism can help us define a meaningful bounded coefficient 
characterizing the efficiency of our heat pump. Indeed, the feedback not only injects energy into 
the system but also an information flow $\dot{\C F}_S$. Using the first law 
(\ref{eq first law with feedback}) at steady state we can rewrite the second law 
(\ref{eq entropy prod feedbackSS}) as 
\begin{equation}
T_R \dot S_{\bf i} = \dot Q^L \eta_C + \dot{\C F}_E + T_R \dot{\C F}_S,
\end{equation}
where $\dot{\C F}_E + T_R\dot{\C F}_S$ can be interpreted as a ``free energy'' injected by the feedback.
To operate as a heat pump, this term has to be positive since $-\dot Q^L>0$. 
As a result by defining the coefficient of performance as 
\begin{equation}\label{eq cop with Feed}
 \kappa_C \equiv \frac{-\dot Q^L}{\dot{\C F}_E + T_R\dot{\C F}_S} = \frac{1}{\eta_C}\left(1-\frac{T_R\dot S_{\bf i}}{\dot{\C F}_E + T_R\dot{\C F}_S}\right),
\end{equation}
the positivity of the entropy production implies the following upper-bounded 
\begin{equation}
 \kappa_C \le \frac{1}{\eta_C}.
\end{equation}
When $\dot{\C F}_S=0$, the feedback plays the role of a pure work source and $\kappa_C = \tilde\kappa$.   

\begin{figure}
\includegraphics[width=0.43\textwidth,clip=true]{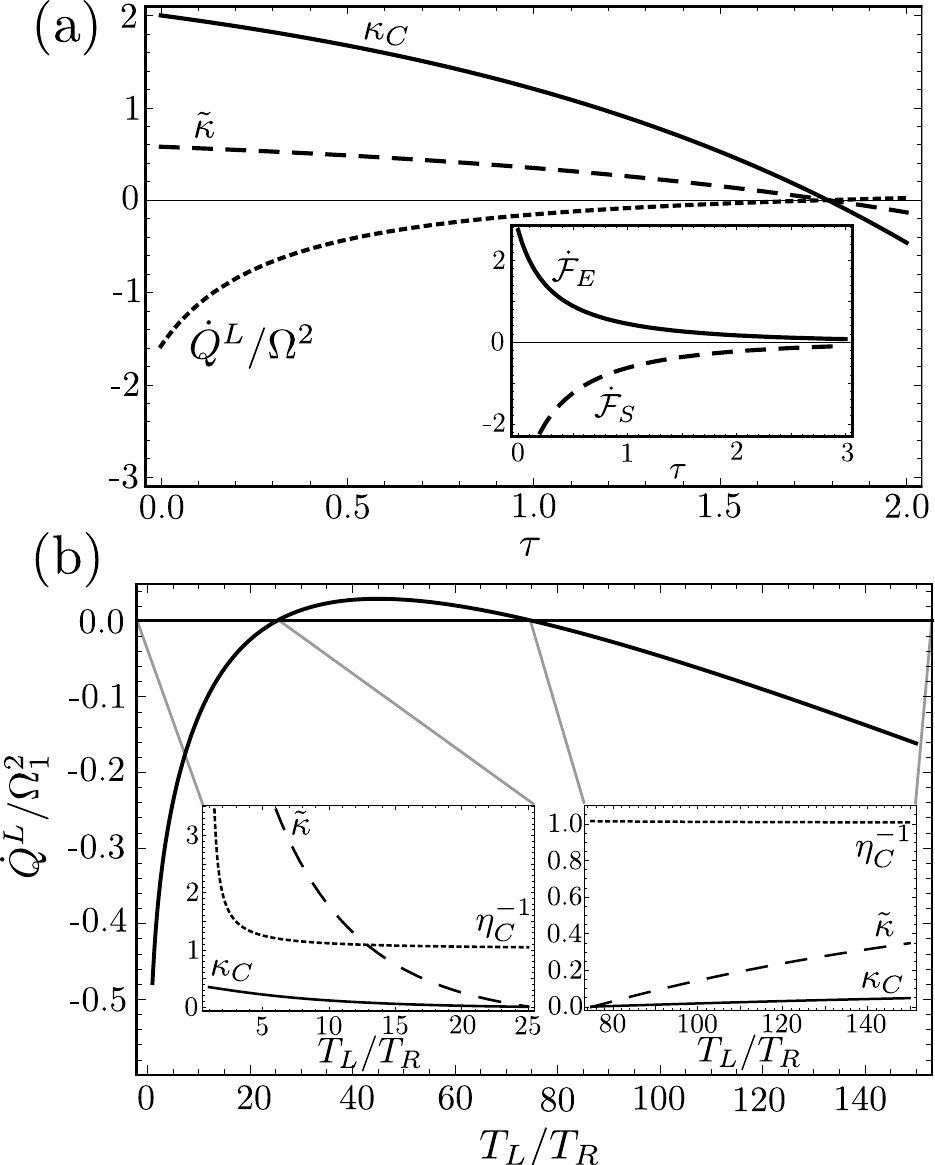}
\caption{\label{fig cop} 
Figure (a) (qubit model): 
$\kappa_C$, Eq. (\ref{eq cop with Feed}), (solid line) and  $\kappa$, Eq. (\ref{eq cop}), (dashed line) as a function 
of the delay time $\tau$ for $T_L = 1$ and $T_R = 1/2$ (thus, $\eta_C^{-1}=2$). The dotted line represents $\dot Q_L$. 
As the feedback parameters we choose $\alpha_+^\nu=0$ and $\alpha_-^\nu=\pi/2$. 
Inset: Energy and entropy injected by the feedback, $\dot{\C F}_E$ (solid line) and $\dot{\C F}_S$ (dashed line) 
as a function of the delay time $\tau$. We also choose $\Omega =\Gamma_L = \Gamma_R = 1$. 
Figure (b) (qutrit model, see appendix \ref{sec qutrit}): No time delay $\tau=0$. Plot of $\dot Q^L$ as a function of $T_L$, 
for $T_R=1$ and $\alpha = \pi/3$. 
Insets: Plot of $\kappa_C$ (solid), $\kappa$ (dashed) and $\eta_C^{-1}$ (dotted) in the regions where $\dot Q^L<0$. 
We choose $\Gamma_L^1 = \Gamma_R^2 = 0.1, \Gamma_L^2 = \Gamma_R^1 = 2, \Gamma_L^\Delta = \Gamma_R^\Delta = 0.01, \Omega_2 = 1.1, \Omega_1 = 1.0$.}
\end{figure}

In Fig. \ref{fig cop}(a), we compare $\kappa_C$ with $\tilde\kappa$ as a function of the delay time for the qubit model 
described in section \ref{qubitModelFeed}. For simplicity we choose the same time delay $\tau$ for all jump types.
The qubit operates as a heat pump for positive $\kappa_C$ (i.e. negative $\dot Q^L$).
The inset shows that for large time delays $\tau \rightarrow \infty$, the effect of the feedback disappears 
and the energy and entropy contribution of the feedback vanish $\dot{\C F}_{E,S} \rightarrow 0$.

A maximum amount of heat is delivered to the left reservoir when $\dot Q^L = \Omega(\gamma_L p_0 - \ov\gamma_L p_1)$ 
is minimized, i.e., when $p_0 \to 0$ and thus $p_1 \to 1$. This can be achieved by choosing a feedback such that 
$\alpha_-^\nu \rightarrow \pi/2, \alpha_+^\nu \rightarrow 0$ for $\nu\in\{L,R\}$ [see (\ref{eq p0 qubit})]. 
In this limit one can even show that the zero delay time limit leads to a vanishing entropy production, 
$\lim_{\tau \rightarrow 0} \dot S_{\bf i} = 0$. Indeed, we can see on Fig. \ref{fig cop}(a) that 
in this reversible limit $\kappa_C$ is maximized and reaches it upper bound $1/\eta_C$. 

In Fig. \ref{fig cop}(b), we use the qutrit model described in appendix \ref{sec qutrit} to show that $\tilde\kappa$ can be larger 
than $\eta_C^{-1}$, because this never happens in the qubit model. The insets compare $\kappa_C$ with $\tilde\kappa$ and $\eta_C^{-1}$
as a function of the temperature of the hot reservoir in the regions where the qutrit operates as a heat pump. 

\subsection{Stabilization of Pure States}
\label{sec stabilization}

The feedback is able to generate steady-state coherences (see Eq. (\ref{eq rho01 qubit})). 
This raises the question whether it is possible to stabilize a \emph{pure} quantum state.
Thus, we are looking for solutions of the equation $0 = \C W^C \rho$ where $\rho$ satisfies 
$\mbox{tr}[\rho^2] = \mbox{tr}[\rho] = 1$ or equivalently we have $\rho \leftrightarrow \varrho = |\psi\rangle\langle\psi|$ 
where $|\psi\rangle$ is a wavefunction.

Immediately after a quantum jump, the qubit finds itself in an energy eigenstate of the system (which is pure). 
After the jump, the feedback rotates the qubit into another pure state which in general involves arbitrary superpositions 
of the energy eigenstates. We denote them as 
\begin{equation}
 \C C^\nu_+\C J_+^\nu |1\rrangle \equiv \rho_+^\nu, ~~~ \C C^\nu_-\C J_-^\nu |0\rrangle \equiv \rho_-^\nu.
\end{equation}
For simplicity, we tune the feedback parameters $\{\alpha_\pm^\nu\}$ so that the state 
of the system right after the control operation is the same independently of the jump 
triggering the control operation, i.e., we demand $\rho_+^L=\rho_+^R=\rho_-^L=\rho_-^R$. 
By choosing $\alpha_+^\nu \equiv \alpha$ and $\alpha_-^\nu \equiv \alpha + \frac{\pi}{2}$ we achieve this
and the state right after the control operation is (in the ordered eigenbasis $\{|1\rangle,|0\rangle\}$)
\begin{equation}\label{target}
 \varrho_{target}(\alpha) = \left(\begin{array}{cc}
                                   \cos^2\alpha &       i\cos\alpha\sin\alpha   \\
                                   -i\cos\alpha\sin\alpha       &       \sin^2\alpha    \\
                                  \end{array}\right).
\end{equation}

We start by considering the vanishing time delay limit.
The system evolution between the jumps is described by the generator $\C W_0$ and in general 
destroys coherences as well as the states purity. A measure of how far the steady state   
is from $\varrho_{target}$ is the \emph{trace distance}. For two arbitrary
density matrices $\varrho_1$ and $\varrho_2$ it is defined by
\begin{equation}
 \mathfrak{D}[\varrho_1,\varrho_2] \equiv \frac{1}{2}\mbox{tr}\sqrt{(\varrho_1-\varrho_2)^2} 
= \frac{1}{2} \sum_i |\lambda_i|,
\end{equation}
where $\lambda_i$ are the eigenvalues of $\varrho_1-\varrho_2$. 
We have $\mathfrak{D}[\varrho_1,\varrho_2] \in [0,1]$. If $\mathfrak{D}[\varrho_1,\varrho_2]=0$ the states 
$\varrho_1$ and $\varrho_2$ are experimentally indistinguishable. 

\begin{figure}
\includegraphics[width=0.48\textwidth,clip=true]{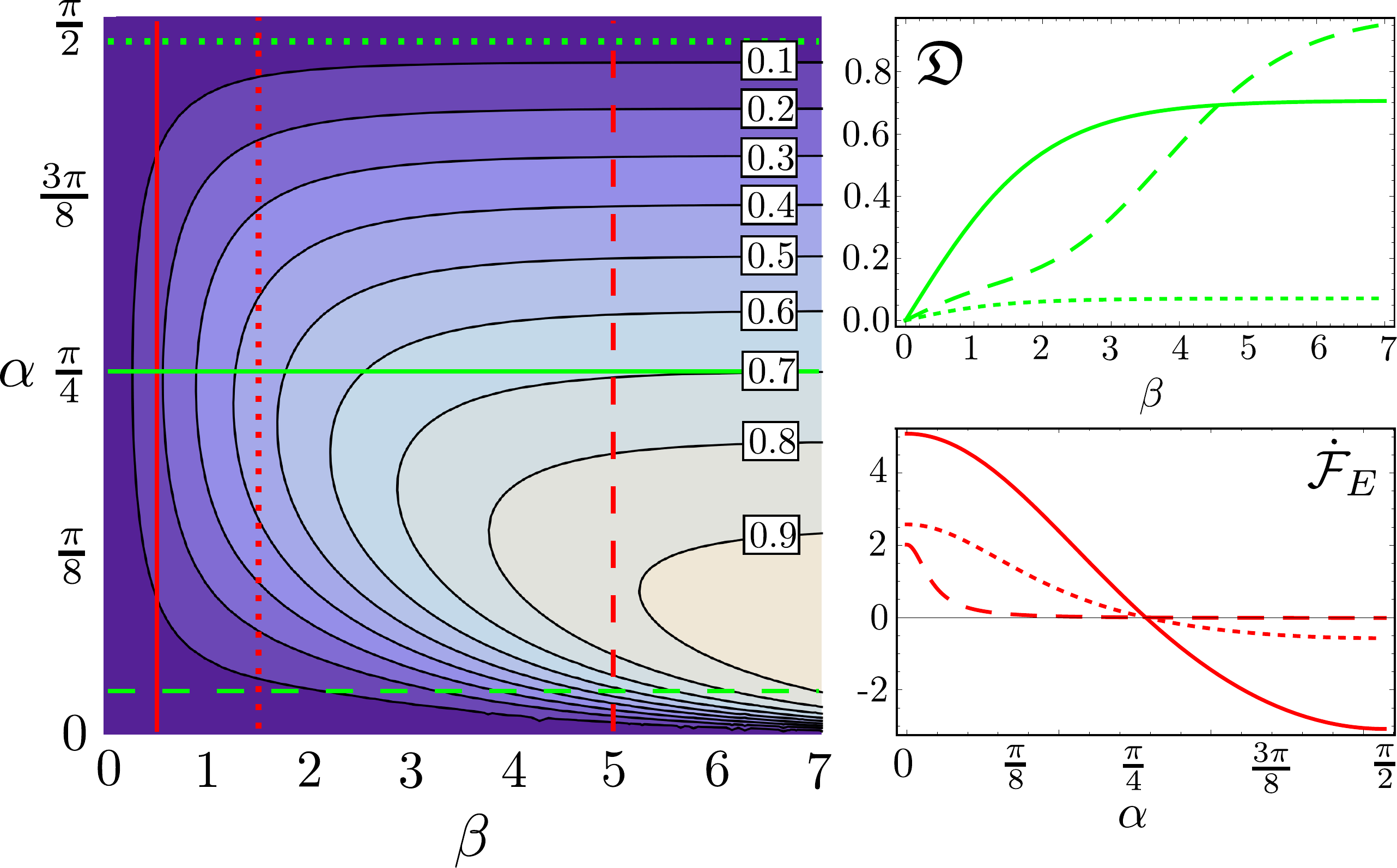}
\caption{\label{fig distance} (Color online.) Contour plot of the distance
$\mathfrak{D}[\varrho_{target},\varrho]$ for varying $\beta$ and $\alpha$
(left side). The horizontal (green) lines corresponds to a plot of the
distance for $\alpha = \pi/20$ (dashed),
$\pi/4$ (solid) and $3/2$ (dotted). The vertical (red) lines corresponds
to plots of the energy injection rate due to the
feedback for $\beta=0.5$ (solid), $1.5$ (dotted) and $5$ (dashed).
Further values were choosen as $\Gamma_L = \Gamma_R = 1$ and $\Omega=1$.}
\end{figure}

As we increase the frequency of the jumps (keeping a zero time delay), the absolute 
magnitude of the terms in $\C W_0$ will increase but $\C W_0$ will also have less 
time to act. This happens for large temperatures when the Bose distributions 
become very large. The resulting effect on $\mathfrak{D}[\varrho_{target},\varrho]$
is explored numerically on Fig. \ref{fig distance} for $\beta_L = \beta_R \equiv \beta$.
The results for $\beta_L \neq \beta_R$ are not qualitatively different. As we see, for large 
temperatures we get closer to $\varrho_{target}$. In fact, it is even possible to show analytically that 
\begin{equation}
 \lim_{\beta_L,\beta_R\rightarrow0} \mathfrak{D}(\beta_L,\beta_R,\alpha) = 0,
\end{equation}
which implies that the steady state of the system coincides with $\varrho_{target}$ for any $\alpha$. In this limit, however, 
also the rate of feedback operations diverges.

\begin{figure}
\includegraphics[width=0.40\textwidth,clip=true]{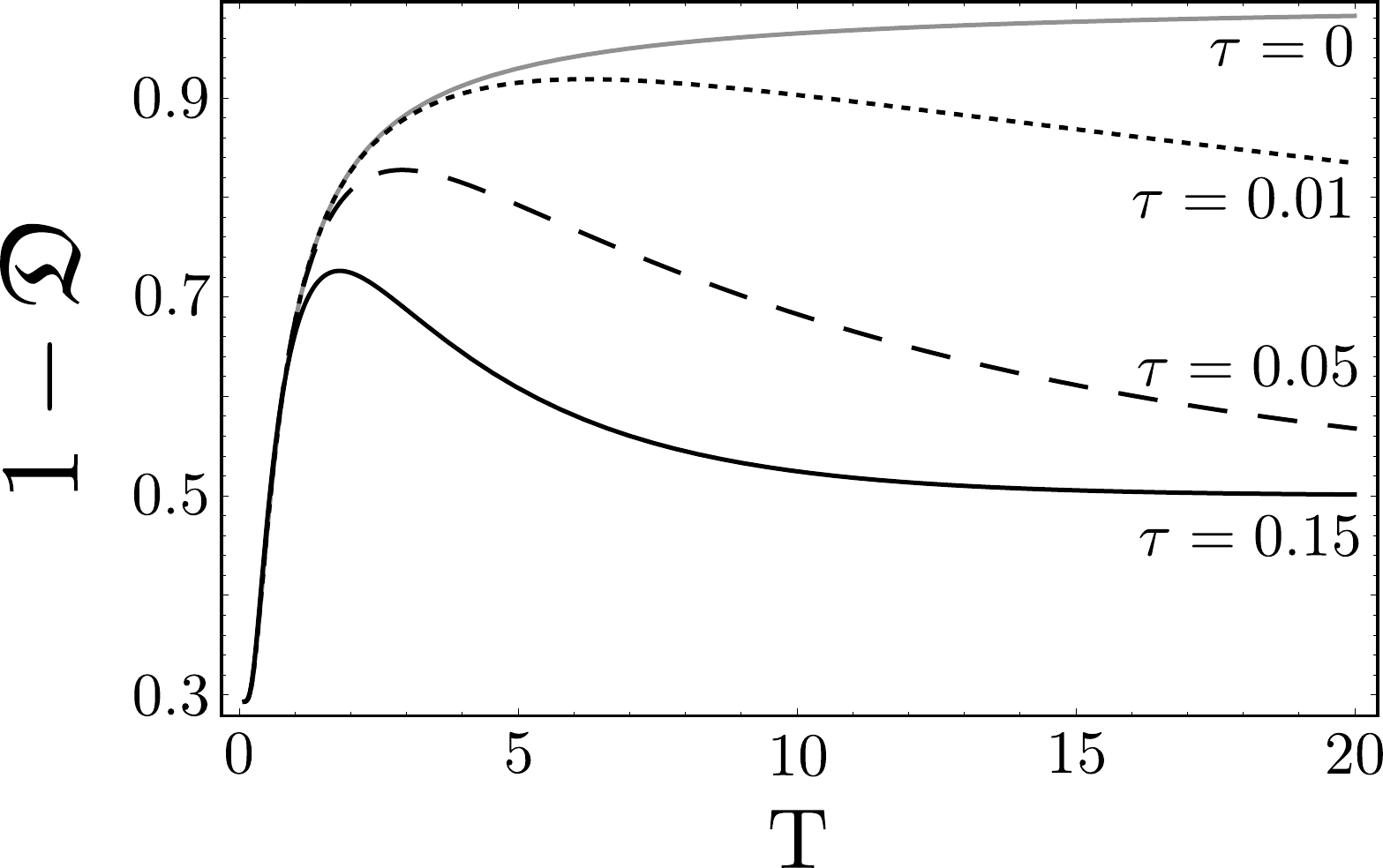}
\caption{\label{fig stab delay} Plot of
$1-\mathfrak{D}[\varrho_{target},\varrho]$ for $\alpha = \pi/4$
over the common reservoir temperature $T$.
All the other parameters are choosen as in Fig. \ref{fig distance}.}
\end{figure}

We now turn to finite delay times. 
In Fig. \ref{fig stab delay} we consider one minus the trace distance 
for the special case $\alpha = \pi/4$ which corresponds to the state $\varrho_{target}(\pi/4) = |\psi\rangle\langle\psi|$ 
with wavefunction $|\psi\rangle = \frac{e^{i\varphi}}{\sqrt{2}}(|1\rangle-i|0\rangle)$.
As expected, when the time between two subsequent jumps becomes smaller, the influence 
of the delay becomes stronger. Thus, for finite delay times we observe the appearance of an 
optimal temperature which maximizes $1-\mathfrak D$, i.e., it minimizes the distance between the stationary state and 
the target state. 

At finite temperature, $\mathfrak{D}[\varrho_{target},\varrho]$ strictly 
vanishes only for $\alpha=0$ and $\alpha=\frac{\pi}{2}$, see Fig. \ref{fig distance} again. This means that 
one can only fully stabilize the excited state $|1\rangle$ and the ground 
state $|0\rangle$. 
To see this theoretically we split the generator as $\C W^C = \C W_0 
+ \C J^C$ where $\C J^C = \sum_\nu (\C C_+^\nu\C J_+^\nu + \C C_-^\nu\C J_-^\nu)$ 
describes the quantum jumps followed by the control operations.
The time evolution between the jumps is therefore governed by $\C W_0$
which, following \cite{BrandesPoltl11, BrandesKiesslich12}, can be 
expressed by a generalized commutator as 
$\C W_0 \rho \leftrightarrow -i\big(\tilde H \varrho - \varrho \tilde H^\dagger\big)$,
where we introduced an effective non-hermitian Hamiltonian
\begin{equation}\label{eq non hermitian hamiltonian}
\tilde H = \frac{1}{2}\left[\big(\Omega-i\sum_\nu\ov\gamma_\nu\big)|1\rangle\langle1|
          + \big(-\Omega-i\sum_\nu\gamma_\nu\big)|0\rangle\langle0|\right].
\end{equation}
In our example, the Liouville space eigenstates of $\tilde H$ and $\tilde H^\dagger$ 
are the same and correspond to the Hilbert space eigenstates of the 
qubit $|0\rangle$ and $|1\rangle$. This means that beside these two states, 
any other state will be destabilized during the evolution between the jumps.

\begin{figure}
\hspace{-0.6cm}
\includegraphics[width=0.51\textwidth,clip=true]{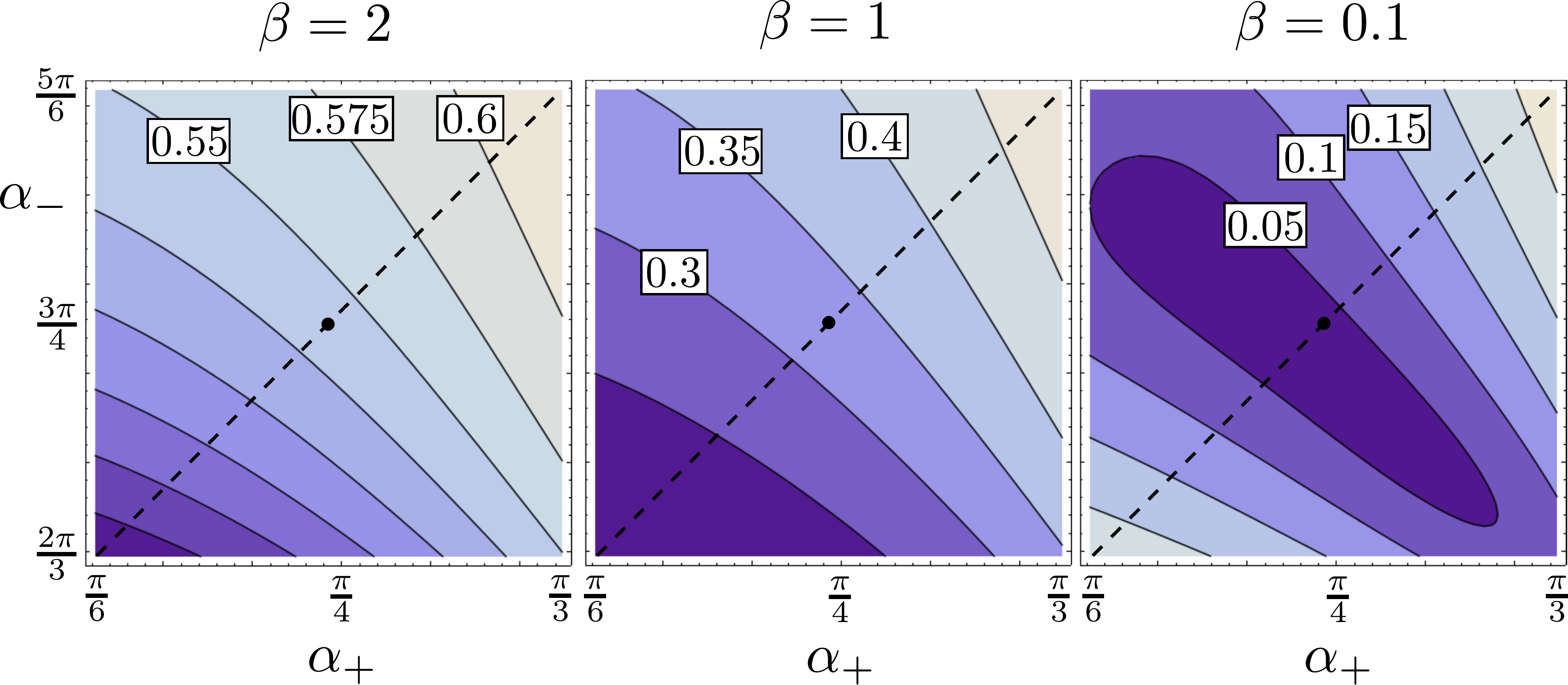}
\caption{\label{fig diff alpha} Contour plot of the distance $\mathfrak{D}[\varrho,\varrho_{target}(\pi/4)]$ over three different 
bath temperatures $\beta_L=\beta_R=\beta$ where the steady state $\varrho$ is obtained for varying $\alpha_+$ and $\alpha_-$. 
The dashed diagonal line corresponds to the previous setting $\alpha_+=\alpha$ and $\alpha_-=\alpha+\pi/2$ and the dot marks the case 
$\alpha=\pi/4$.}
\end{figure}

Until now we tuned the feedback parameters such that the state of the system 
after any control operation is always the same. We now relax this assumption
by considering $\mathfrak{D}[\varrho_{target},\varrho]$ as a function of 
$\alpha_+^\nu\equiv\alpha_+$ and $\alpha_{-}^\nu\equiv\alpha_-$ where the target 
state is the pure state $\varrho_{target}(\pi/4)$, which corresponds to an equally 
weighted superposition of the excited state and the ground state. 
On Fig. \ref{fig diff alpha} we see that -- due to the incoherent time evolution 
between the jumps -- it is better to choose slightly different values from 
$\alpha_+ = \frac{\pi}{4}$ and $\alpha_- = \frac{3\pi}{4}$ considered before. 
This discrepancy vanishes for large temperatures. 

We finally note that in this section on steady state coherence stabilization,
the entropy production (\ref{eq entropy prod qubit feedback}) always vanished 
since $\beta_L=\beta_R$ and $f_L=f_R$. This shows that the notion of reversibility 
in presence of feedback resulting from our treatment clearly does not imply 
canonical steady states and even allows for pure states. 
Furthermore, one may even extract energy from the system by stabilizing a certain set of states, as indicated in Fig. \ref{fig distance}. 

\section{Conclusions} \label{conc}

We considered open quantum systems in contact with multiple reservoirs and subjected to quantum feedback operations 
triggered by the detection of transfer processes between the system and its reservoirs. The feedback operations are 
instantaneous unitary operations in the system Hilbert space, which can be performed immediately after the detection 
or after a finite time delay.  

We showed that the dynamics of such systems can still be analyzed within the framework of stochastic thermodynamics
despite the fact that the quantum feedback operation can stabilize coherences in the stationary states which would 
vanish in absence of feedback. The quantum feedback operation injects energy and entropy (or information) into the
system and thus modifies the energy balance (the first law) as well as the entropy balance (the second law). 

In absence of time delay, the effect of the feedback on the thermodynamic description of the system can be 
understood classically. Each time a monitored reservoir-induced transition occurs in the system, the system 
ends up in one of its energy eigenstate. The effect of the feedback is to induce a transition between this 
eigenstate and the other ones with a given transition probability. This results in a change of the energy 
as well as entropy of the system.      

We applied the formalism to study a qubit in contact with two reservoirs and operating as a heat pump.
We showed that due to the feedback, the coefficient of performance used to characterize the efficiency 
of conventional heat pumps is not bounded by the inverse Carnot efficiency anymore. We proposed a new 
definition of the coefficient of performance that is bounded by our theory. We also analyzed the effect 
of time delay on the heat pump operation. Finally, we demonstrated that the quantum feedback operation 
can be used to stabilize coherences in nonequilibrium steady states including pure states. 

\section*{Acknowledgments} 

Useful discussions with Victor Bastidas are acknowledged. 
Financial support by the DFG (SCHA 1646/2-1, SFB 910, and GRK 1558) and the 
National Research Fund, Luxembourg (project FNR/A11/02) is gratefully acknowledged.


\appendix

\section{Full Counting Statistics}
\label{sec full counting statistics}

We derive the currents (\ref{eq steady state current formula}) and (\ref{eq probability flow with feedback}) 
using full counting statistics methods \cite{EspositoReview}. 
We assign a counting field $\chi_{(i,j)}^\nu$ to every transition from system state $j$ to $i$ due to the reservoir $\nu$. 
The corresponding jump operators are denoted $\C J_{j\rightarrow i}^\nu$ and $\ov{\C J}_{i\rightarrow j}^\nu$. 
The generator then takes the form 
\begin{equation}\label{eq generator with counting fields}
 \C W(\boldsymbol\chi) = 
\C W_0 + \sum_{\nu} \sum_{i>j} \big(e^{i\chi_{(i,j)}^\nu}\C J_{j\rightarrow i}^\nu + e^{-i\chi_{(i,j)}^\nu}\ov{\C J}_{i\rightarrow j}^\nu\big).
\end{equation}
If $\boldsymbol\chi$ denotes the vector of all counting fields, the formal solution of the density 
matrix evolved with this generator is 
\begin{equation}\label{FormalSolCS}
\rho(\boldsymbol\chi,t) = e^{\C W(\boldsymbol\chi)t} \rho(0).
\end{equation}
It turns out that $\mbox{tr}[\rho(\boldsymbol\chi,t)]$ is the moment generating function associated 
to the net integrated probability currents between the states. Thus, the average net integrated current 
associated to the transition $j \to i$ and due to reservoir $\nu$ is calculated by taking the 
derivative of the moment generating function with respect to the counting field $\chi_{(i,j)}^\nu$
\begin{equation}
 \langle n\rangle_{(i,j)}^\nu(t) = \left.\frac{\partial}{\partial(i\chi_{(i,j)}^\nu)}\mbox{tr}[\rho(\boldsymbol\chi,t)]\right|_{\boldsymbol\chi=0}.
\end{equation}
The time derivative yields the current 
\begin{equation}
 I_{(i,j)}^\nu(t) = \frac{\partial}{\partial t}\langle n\rangle_{(i,j)}^\nu(t).
\end{equation}
Applying the time derivative to (\ref{FormalSolCS}), we obtain 
\begin{equation}
 \begin{split}
  I(t)	&=\left.\frac{\partial}{\partial(i\chi_{(i,j)}^\nu)}\mbox{tr}\big[\C W(\boldsymbol\chi)e^{\C W(\boldsymbol\chi)t}\rho(0)\big]\right|_{\boldsymbol\chi=0} \\
	&=\mbox{tr}\big[\C W'(0)e^{\C W(0)t}\rho(0) + \C W(0) \big(e^{\C W(0)t}\big)' \rho(0)\big],
 \end{split}
\end{equation}
where $\C W'(0)$ and $(e^{\C W(0)t})'$ are shorthand notations for the derivative
with respect to $(i\chi_{(i,j)}^\nu)$ evaluated at $\boldsymbol\chi=0$. 
Since the second term vanishes due to the fact that the generator is norm preserving, $\sum_i\C W_{ij} = 0$, 
and using $\rho(t) = e^{\C W(0)t}\rho(0)$ we find that
\begin{equation}
 I_{(i,j)}^\nu(t) = \mbox{tr}[\C W'(0)\rho(t)].
\end{equation}
Using the form of the generator (\ref{eq generator with counting fields}) yields to the 
desired expression (\ref{eq steady state current formula}). 

Reproducing this argument in presence of feedback leads to
\begin{equation}
 I_{(i,j)}^\nu(t) = \mbox{tr}\left[\big(\C C_{j\rightarrow i}^\nu\C J_{j\rightarrow i}^\nu-\ov{\C C}_{i\rightarrow j}^\nu\ov{\C J}_{i\rightarrow j}^\nu\big)\rho(t)\right].
\end{equation}
Evaluating the trace by using the explicit form of the jump operators and (\ref{eq useful formula}) 
gives after some straightforward calculation (\ref{eq probability flow with feedback}). 

\begin{widetext}
\section{Derivation of the First and Second Law with and without Feedback Control}
\label{sec derivation first law}

The time derivative of the average energy of a system described by a rate equation $\dot p_i(t) = \sum_j \C W_{ij}p_j(t)$ with $\C W = \sum_\nu \C W^{(\nu)}$ reads
\begin{equation}\label{eq derivation first law}
 \dot E(t) = \sum_{i} E_i \dot p_i(t) \overset{(\star)}{=} \sum_\nu \sum_{i,j} (E_i-E_j) \C W^{(\nu)}_{ij} p_j(t) 
= \sum_\nu \sum_{i>j} (E_i-E_j) \big[\C W^{(\nu)}_{ij}p_j(t)-\C W^{(\nu)}_{ji}p_i(t)\big].
\end{equation}
For step $(\star)$, we used the fact that the rate equation preserves probability: $\sum_i \C W_{ij} = 0$ for every $j \in \{1,\dots,M\}$. 
Using the definition of the heat flow (\ref{eq heat flow in general}), Eq. (\ref{eq derivation first law}) immediately gives the first law 
of thermodynamics without feedback (\ref{eq first law}). In presence of feedback we have to insert the modified population generator from 
Eq. (\ref{eq pop generator with feedback}). This yields after some calculations to the expression 
\begin{equation}
 \begin{split}
  \dot E(t)	&=	\sum_\nu \sum_{i>j} (E_i-E_j)(\gamma_{j\rightarrow i} p_j(t) - \ov\gamma_{i\rightarrow j} p_i(t))	\\
		&+	\sum_\nu  \sum_{i>j} \left\{\big[\sum_k (\C C_{j\rightarrow i}^\nu)_{ki} E_k - E_i\big] \gamma_{j\rightarrow i}^\nu p_j(t) + \big[\sum_k (\ov{\C C}_{j\rightarrow i}^\nu)_{kj} E_k - E_j\big] \ov\gamma_{i\rightarrow j}^\nu p_i(t)\right\}.
 \end{split}
\end{equation}
The first term equals again the sum over all heat flows and the rest equals the rate of energy injection due to the feedback 
$\dot{\C F}_E(t) = \sum_\nu \dot{\C F}_E^{(\nu)}(t)$. This can be confirmed by evaluating Eq. (\ref{eq work feedback general}). 
Thus, we end up with the first law stated in Eq. (\ref{eq first law with feedback}). 

We now turn to the second law of thermodynamics. For this we want to calculate the information flow at steady state. 
The entropy flow, Eq. (\ref{eq entropy flow general}), can be written after some algebra as 
\begin{equation}
 \dot S_{\bf e}(t) = \sum_\nu  \sum_{i>j} \big[(\C W^C)^{(\nu)}_{ij} p_j(t) - (\C W^C)^{(\nu)}_{ji} p_i(t)\big] \ln\frac{(\C W^C)^{(\nu)}_{ji}}{(\C W^C)^{(\nu)}_{ij}}.
\end{equation}
In absence of feedback, using local detailed balance (\ref{eq local detailed balance}) and the definition of the heat flow 
(\ref{eq heat flow in general}), we get the second law of thermodynamics stated in Eq. (\ref{eq second law no feedback}) 
with $\dot S_{\bf e}(t) = \sum_\nu \dot Q^{(\nu)}(t)/T_\nu$. 
In presence of feedback, we have to use the modified population generator from Eq. (\ref{eq pop generator with feedback}). 
After separating the heat flows, we have
\begin{equation}
 \begin{split}
  \C F_S^{(\nu)}(t) - \frac{\dot Q^{(\nu)}(t)}{T_\nu}   &=      -\sum_{m>m'}
\left(\sum_{i|i>m'} \gamma_{m'\rightarrow i}^\nu (\C
C^\nu_{m'\rightarrow i})_{mi} + \sum_{j|j<m'} \ov\gamma_{m'\ra j}^\nu
(\ov{\C C}_{m'\ra j}^\nu)_{mj}\right) p_{m'}(t) \ln\frac{(\C
W^C)^{(\nu)}_{m'm}}{(\C W^C)^{(\nu)}_{mm'}}     \\
                                        &+      \sum_{m>m'} \left(\sum_{i|i>m} \gamma_{m\rightarrow i}^\nu (\C
C^\nu_{m\rightarrow i})_{m'i} + \sum_{j|j<m} \ov\gamma_{m\ra j}^\nu
(\ov{\C C}_{m\ra j}^\nu)_{m'j}\right) p_{m}(t) \ln\frac{(\C
W^C)^{(\nu)}_{m'm}}{(\C W^C)^{(\nu)}_{mm'}}.
 \end{split}
\end{equation}

where $\sum_{i|i>m}$ denotes a sum running over those $i$ which fulfill $i>m$. 

\section{Steady state of the Qubit}
\label{sec steady state qubit}

For completeness we give the exact steady state of the feedback controlled qubit obtained by solving the equation 
$0 = \C W^C\rho$ with the generator (\ref{eq feedback generator qubit}) and (\ref{eq Wcoh qubit}). 
The populations read 
\begin{equation}\label{eq p0 qubit}
 p_0 = 1-p_1 =\frac{\ov\gamma_L\cos^2\alpha_-^L + \ov\gamma_R\cos^2\alpha_-^R}{\gamma_L\cos^2\alpha_+^L + \ov\gamma_L\cos^2\alpha_-^L + \gamma_R\cos^2\alpha_+^R + \ov\gamma_R\cos^2\alpha_-^R}
\end{equation}
and the coherences are given by 
 \begin{equation}\label{eq rho01 qubit}
  \begin{split}
   \rho_{01} = \rho_{10}^* 
                &= \frac{2i \gamma_L \cos\alpha_+^L \left(\ov\gamma_L\cos\alpha_-^L \sin(\alpha_-^L-\alpha_+^L) +\ov\gamma_R\cos\alpha_-^R \sin(\alpha_-^R-\alpha_+^L)\right)}{\left(-2 i \Omega +\gamma_L+\ov\gamma_L+\gamma_R+\ov\gamma_R\right) \left(\gamma_L\cos^2\alpha_+^L +\ov\gamma_L\cos^2\alpha_-^L +\gamma_R\cos^2\alpha_+^R+\ov\gamma_R-\ov\gamma_R\sin^2\alpha_-^R\right)}	\\
		&+	\frac{2i \gamma_R\cos\alpha_+^R \left(\ov\gamma_L\cos\alpha_-^L \sin(\alpha_-^L-\alpha_+^R)+\ov\gamma_R\cos\alpha_-^R \sin(\alpha_-^R-\alpha_+^R)\right)}{\left(-2 i \Omega +\gamma_L+\ov\gamma_L+\gamma_R+\ov\gamma_R\right) \left(\gamma_L\cos^2\alpha_+^L+\ov\gamma_L\cos^2\alpha_-^L+\gamma_R\cos^2\alpha_+^R+\ov\gamma_R-\ov\gamma_R\sin^2\alpha_-^R\right)}.
  \end{split}
 \end{equation}

\end{widetext}

\section{The qutrit} \label{sec qutrit}

The qutrit model is a three-level system with Hamiltonian 
\begin{equation}
 H_S = \Omega_2|2\rangle\langle2| + \Omega_1|1\rangle\langle1| + 0|0\rangle\langle0|,
\end{equation}
where $\Omega_2>\Omega_1>0$ defines the levels of the qutrit. 
As in the qubit model, the interaction in chosen in the RWA 
\begin{equation}
 V = \sum_{\nu,\bb q} \sum_{i<j} T_{\bb q\nu}(b^\dagger_{\bb q\nu} |i\rangle\langle j| + b_{\bb q\nu}|j\rangle\langle i|).
\end{equation}
The qutrit is coupled to a left and a right reservoir. The population generator in the ordered basis $(p_0,p_1,p_2)$ reads
\begin{equation}
 \C W_{pop} = \sum_\nu\left(\begin{array}{ccc}
			     -\gamma_\nu^1-\gamma_\nu^2	&	\ov\gamma_\nu^1				&	\ov\gamma_\nu^2				\\
			     \gamma_\nu^1		&	-\gamma_\nu^\Delta-\ov\gamma_\nu^1	&	\ov\gamma_\nu^\Delta			\\
			     \gamma_\nu^2		&	\gamma_\nu^\Delta			&	-\ov\gamma_\nu^2-\ov\gamma_\nu^\Delta	\\
			    \end{array}\right).
\end{equation}
As in the qubit model, the rates are expressed in terms of the Bose distribution 
$\gamma_\nu^\omega = \Gamma_\nu^\omega n_\nu(\omega), \ov\gamma_\nu^\omega = \Gamma_\nu^\omega (1+n_\nu(\omega))$, 
where $\omega$ denotes an energy difference between system states and we abbreviated 
$\gamma_\nu^1 \equiv \gamma_\nu^{\Omega_1}, \gamma_\nu^2 \equiv \gamma_\nu^{\Omega_2}, \gamma_\nu^\Delta \equiv \gamma_\nu^{\Omega_2-\Omega_1}$. 


We consider the following control scheme. Whenever the transition $|0\rangle\rightarrow|1\rangle$ is detected 
we apply the control operation $U = \exp[i\alpha(|1\rangle\langle2|+|2\rangle\langle1|)]$. This rotates the 
level $|1\rangle$ to the superposition $\cos\alpha|1\rangle+i\sin\alpha|2\rangle$. 
The resulting population generator reads 
\begin{equation}\label{eq pop generator qutrit}
 \C W^C_{pop} = \sum_\nu\left(\begin{array}{ccc}
			     -\gamma_\nu^1-\gamma_\nu^2			&	\ov\gamma_\nu^1				&	\ov\gamma_\nu^2				\\
			     \gamma_\nu^1\cos^2\alpha			&	-\gamma_\nu^\Delta-\ov\gamma_\nu^1	&	\ov\gamma_\nu^\Delta			\\
			     \gamma_\nu^1\sin^2\alpha+\gamma_\nu^2	&	\gamma_\nu^\Delta			&	-\ov\gamma_\nu^2-\ov\gamma_\nu^\Delta	\\
			    \end{array}\right).
\end{equation}


\end{document}